\documentstyle[12pt,emulateapj,psfig]{article}

\newcommand\sz     {$S_{\rm z}$}

\newcommand\pp     {$\pm$}
\newcommand\pers     {s$^{-1}$}
\newcommand\micros  {$\mu$s}

\righthead{Timing XTE J1806--246}
\slugcomment{Submitted to ApJ, Feb 1998}

\begin{document}

\title{The discovery of a 7--14 Hz Quasi-Periodic Oscillation in the
X-ray Transient XTE J1806--246}

\author{Rudy Wijnands \& Michiel van der Klis}

\affil{Astronomical Institute ``Anton Pannekoek'',
University of Amsterdam, and Center for High Energy Astrophysics,
Kruislaan 403, NL-1098 SJ Amsterdam, The Netherlands;
rudy@astro.uva.nl, michiel@astro.uva.nl}

\begin{abstract}
We have studied the correlated X-ray spectral and X-ray timing
behavior of the X-ray transient XTE J1806--246 using data obtained
with the proportional counter array onboard the {\it Rossi X-ray
Timing Explorer}. In the X-ray color-color diagram two distinct
patterns are traced out. The first pattern is a curved branch, which
is observed during the rise and the decay of the outburst. This
pattern resembles the so-called banana branch of those low-luminosity
neutron star low-mass X-ray binaries (LMXBs) which are referred to as
atoll sources. The power spectrum of XTE J1806--246 on this curved
branch consisted of a power law and a cutoff power law component. The
presence of these components and their dependence on position of the
source on the branch is also identical to the behavior of atoll
sources on the banana branch.  Near the end of its outburst, XTE
J1806--246 formed patches in the color-color diagram, the spectrum was
harder, and the power spectrum showed strong band limited noise,
characteristic of the atoll sources in the island state.  A second
pattern was traced out during the only observation at the peak of the
outburst. It consists of a structure which we interpret as formed by
two distinct branches. This pattern resembles the normal-flaring
branches of the high-luminosity neutron star LMXBs (the Z
sources). The discovery of a 7--14 Hz QPO during this observation
strengthens this similarity.  We conclude that if XTE J1806--246 is a
neutron star, it is most likely an atoll source that only at the peak
of its outburst reached a luminosity level sufficiently high to show
the type of QPO that Z sources show on their normal and flaring
branches.

\end{abstract}


\section{Introduction \label{intro}}

The new X-ray transient XTE J1806--246 was first detected (Marshall \&
Strohmayer 1998) with the {\it Rossi X-ray Timing Explorer} ({\it
RXTE}) and later most likely identified in the radio (Hjellming,
Mioduszewski, \& Rupen 1998) and optical (Hynes, Roche, \& Haswell
1998) bands.  The source is positionally coincident with the known
X-ray transient 2S 1803--245 (Jernigan et al. 1978) and with the X-ray
burster SAX J1806.8--2435 (Muller et al. 1998). If the latter
identification is correct then this would imply that XTE J1806--246 is
a low-mass X-ray binary (LMXB) harboring a low magnetic field strength
neutron star and not a black hole.

The neutron star LMXBs have been classified into the so-called Z
sources and the atoll sources on the basis of their correlated X-ray
spectral and X-ray timing behavior (Hasinger \& van der Klis
1989). The Z sources trace out a Z shape like track in the X-ray
color-color diagram (CD) with the branches called, from top to bottom,
the horizontal branch (HB), the normal branch (NB), and the flaring
branch (FB). The power spectra of the Z sources show on the HB strong
band-limited noise (called low frequency noise or LFN) with a cutoff
frequency of several Hertz, and, simultaneous with this, QPOs between
15 and 60 Hz, which are called horizontal branch QPOs or HBOs. On the
NB QPOs appear between 5 and 7 Hz which are called normal branch QPOs
or NBOs.  HBOs and NBOs often occur simultaneously.  In several Z
sources, the NBOs smoothly merge with 7--20~Hz QPOs seen on the FB,
the flaring branch QPOs or FBOs.  On all branches two additional noise
components are found, one at very low frequencies (the very low
frequency noise or VLFN), following a power law, and one at
frequencies above 10 Hz (the high frequency noise or HFN), which cuts
off between 50 and 100~Hz.

The atoll sources trace out a curved branch in the CD, which can be
divided in two parts: the island state and the banana branch. When the
source is in the island state motion in the CD is not very fast and
can take several weeks.  The power spectrum is dominated by very
strong (sometimes more than 20\% rms amplitude) band-limited noise.
The band-limited noise is also called HFN but it is at much lower
frequencies than the HFN observed in the Z sources and most likely
they are not related. Instead, it has been proposed that Z source LFN
and atoll source HFN are similar phenomena (van der Klis 1994).  On
the banana branch the source moves much faster through the CD (on time
scales of hours to days) and in the power spectrum only a weak
(several percent rms amplitude) power law noise component at low
frequencies is observed (the VLFN). The part of the banana branch
closest to the island state is called the lower banana branch, the
part away the upper banana branch. From the island state via the lower
to the upper banana branch the VLFN tends to get stronger and steeper
while the HFN becomes much weaker.

The physical parameter governing the state changes in a given source
is nearly certainly the mass accretion rate (see van der Klis 1995 for
a summary of the evidence). Some of the differences between the Z and
the atoll sources are also due to the differences in the average
accretion rate, but it has been proposed that differences in the
neutron star magnetic field strength also play a role (e.g., Hasinger
\& van der Klis 1989; Psaltis, Lamb, \& Miller 1995).

With the launch of {\it RXTE} at the end of 1995, in both the Z
sources and the atoll sources two other types of QPOs were discovered
at much higher frequencies, between 200 and 1200 Hz (the kHz QPOs; see
van der Klis 1999 for a recent review). In the Z sources, these kHz
QPOs are detected on the HB and the upper part of the NB. In the Z
source Sco X-1, the kHz QPOs are detected all the way up to the lower
part of the FB. In atoll sources, the kHz QPOs are strong in the
island state, weaker on the lower part of the banana branch, and
absent on the upper part of the banana branch. The similar properties
of the kHz QPOs in the Z sources and the atoll sources suggest that
these QPOs are most likely due to the same physical mechanism in both
types of sources.

Using data obtained with {\it RXTE} not only kHz QPOs were discovered
in the atoll sources, but also QPOs with frequencies around 60 to 80
Hz (Strohmayer et al. 1996; Wijnands \& van der Klis 1997; Ford et
al. 1997; Homan et al. 1998; Wijnands et al. 1998) and in one atoll
source a QPO was observed near 7 Hz (Wijnands, van der Klis, \&
Rijkhorst 1999). The properties of the 60--80 Hz QPOs suggest that
they could be due to the same physical mechanism as the HBOs in Z
sources (e.g., Homan et al. 1998; Psaltis, Belloni, van der Klis
1999). Wijnands et al. (1999) tentatively suggest that the 7 Hz QPO
can be identified with the NBOs in the Z sources. These recent {\it
RXTE} results show that whatever causes the phenomenological
differences between the Z and the atoll sources, it does not prevent
similar QPO phenomena from occurring (albeit with differences in
incidence and perhaps strength) in both source types.

Here we report on the correlated X-ray spectral and X-ray timing
behavior of XTE J1806--246 during the rise, the peak, and the decay of
its 1998 outburst. We report the discovery of a 7--14 Hz QPO during
the peak of the outburst. A preliminary announcement of this discovery
was already made by Wijnands \& van der Klis (1998). The correlated
X-ray spectral and X-ray timing behavior during the rise and the decay
is consistent with that of an atoll source. The outburst peak
properties are more reminiscent to Z source normal/flaring branch
behavior, which suggests that the lack of phenomena characteristic of
the normal/flaring branch in most atoll sources is due to a difference
in accretion rate only.

When this paper was essentially completed we became aware of a paper
by Revnivtsev, Borozdin, \& Emelyanov (1999) which uses the same
data. They concentrated on the analysis of the 9 Hz QPO and with
respect to that feature obtained results consistent with ours.

\section{Observations and Analysis  \label{observations}}

XTE J1806--246 was observed by the proportional counter array (PCA)
onboard {\it RXTE} as part of public target of opportunity
observations for a total of $\sim$32 ksec (see Table~\ref{obslog} for
a log of the observations). During all observations, data were
obtained in 129 photon energy bands covering the energy range 2--60
keV. Simultaneous data were obtained at different epochs with time
resolutions of 122 \micros~or 16 \micros~in 67 (April 27) or 18 (the
other data) photon energy bands covering the range 2--60 keV. During
all observations except the April 27 one, data were also obtained with
8 ms time resolution in 16 bands covering the range 2--13 keV.

We calculated FFTs using 16-s data segments of the 122 \micros~and 16
\micros~data in order to study the $>$ 100 Hz variability. We also
made 128 s FFTs using the 8 ms and 16 \micros~data in different photon
energy bands in order to study the energy dependence of the QPO and
the noise components.  Finally, we made 128 s cross-spectra between
different energy bands from the same data in order to study the time
lags of the QPO.  To determine the properties of the 7--14 Hz QPO, we
fitted the power density spectra of May 3 with a function containing a
constant (representing the dead-time modified Poisson level), a power
law (representing the underlying continuum), and a Lorentzian
(representing the QPO). To determine the properties of the peaked
noise component, we fitted the power density spectra of the other
data, after subtracting the dead-time modified Poisson level, with a
power law and an exponentially cutoff power law (representing the
peaked noise). The uncertainties in the fit parameters were determined
using $\Delta\chi^2=1$. Upper limits correspond to a 95\% confidence
level.

The PCA light curve, the CDs, and the hardness-intensity diagrams
(HIDs) were created using 64 s averages of the 16 s data.  In the CDs,
the soft color is defined as the logarithm of the
3.5--6.4~keV/2.0--3.5~keV count rate ratio and the hard color as the
logarithm of the 9.7--16.0~keV/6.4--9.7~keV count rate ratio. The
definition of the colors in the HIDs are the same as in the CDs. The
count rates in the HIDs are for the photon energy range 2.0--16.0 keV.
All count rates quoted in this paper are for 5 detectors. The count
rates used in the diagrams are background-subtracted but not dead-time
corrected. The dead-time correction is 3\%--5\%.

\section{Results \label{results}}

\begin{figure}[]
\begin{center}
\begin{tabular}{c}
\psfig{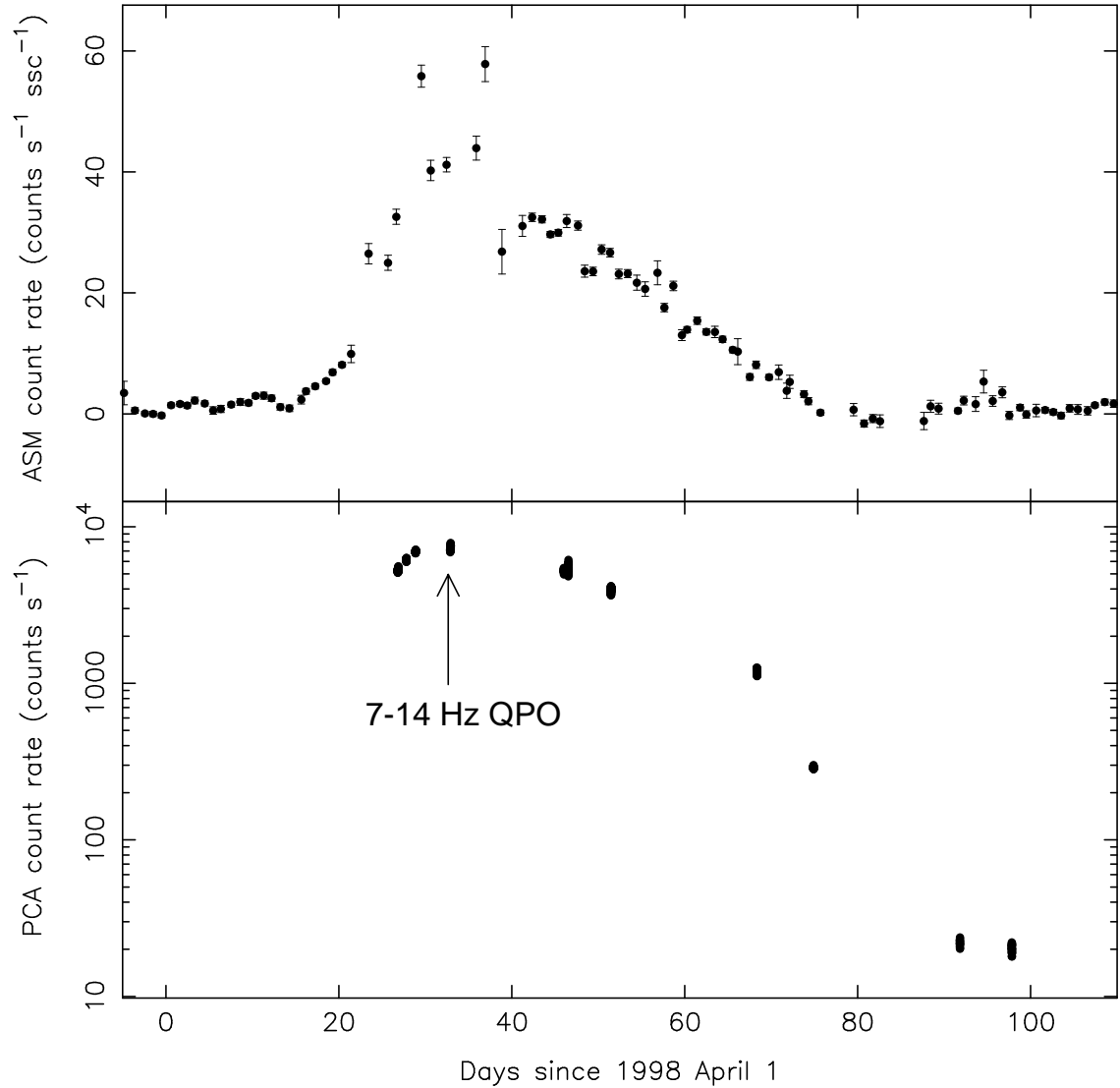}
\end{tabular}
\figcaption{The daily averaged {\it RXTE} All Sky Monitor light curve
(1.3--12.1 keV; {\it top}) and the {\it RXTE}/PCA light curve ({\it
bottom}) of XTE J1806--246. The PCA count rates are in the energy band
2.0--16 keV and are background subtracted but not deadtime
corrected. Each point in the PCA light curve represents 64 s averages.
\label{fig:lightcurves} }
\end{center}
\end{figure}

The outburst light curve is presented in
Figure~\ref{fig:lightcurves}. During 1998 April the source was rising
from $\sim$5000 to $\sim$7200 counts s$^{-1}$ (2.0--16.0 keV). It
reached a maximum of between 7000--8000 counts s$^{-1}$ in the 1998
May 3 observation. After May 3 the count rate decreased; the source
became almost undetectable ($\sim$10--20 counts \pers) by July (see
Table~\ref{obslog}).

\begin{figure}[]
\begin{center}
\begin{tabular}{c}
\psfig{figure=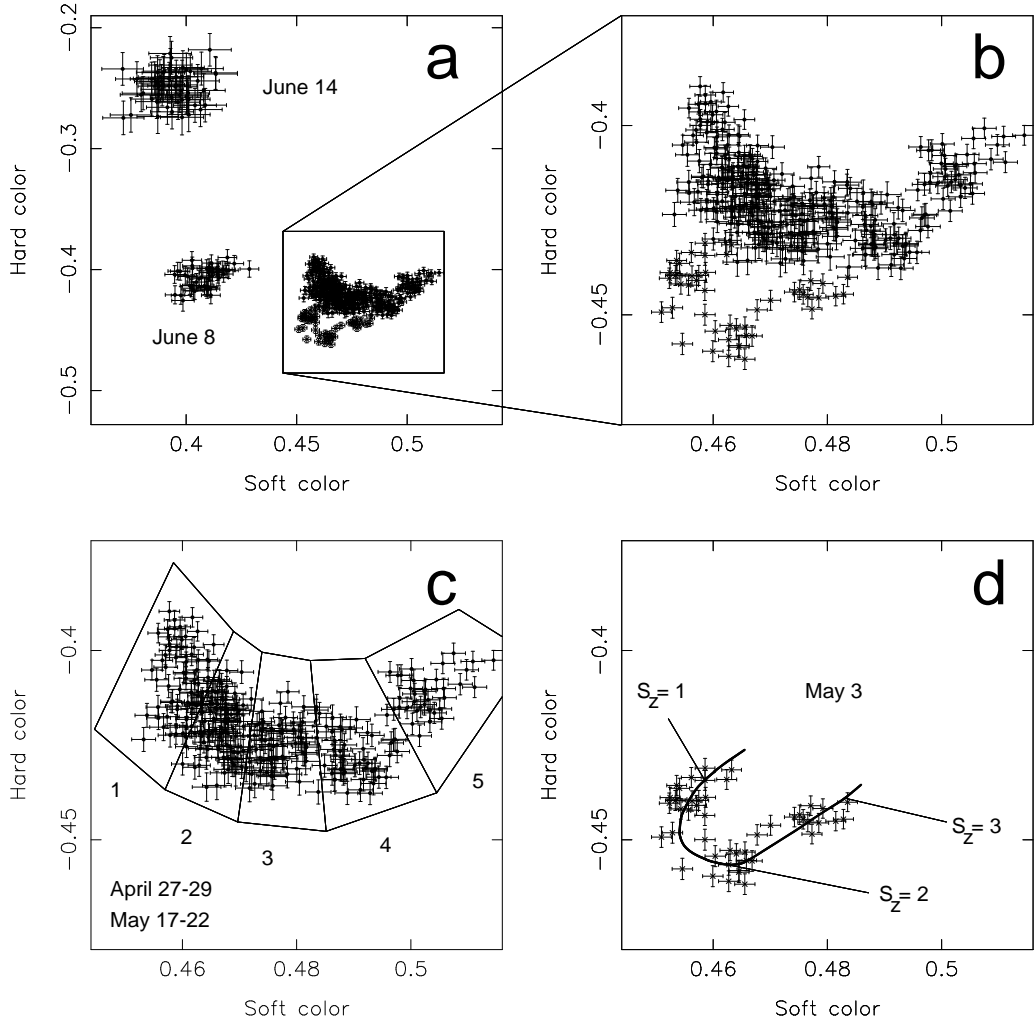,width=14cm}
\end{tabular}
\figcaption{Color-color diagrams of XTE J1806--246 created from the
data obtained between 1998 Apr 27 and Jun 14 ({\it a}). A blow up of
the data obtained between 1998 Apr 27 and May 22 is given in {\it
b}. The rise and the decay data obtained in the same interval as in
{\it b} are given in {\it c}, the data obtained during the peak of the
outburst are given in {\it d}. The soft color is the logarithm of the
count rate ratio between 3.5--6.4 and 2.0--3.5 keV and the hard color
is the logarithm the count rate ratio between 9.7--16.0 and 6.4--9.7
keV. The count rates which are used to calculate the colors are
corrected for background.  In {\it c}, the areas used to select the
power spectra are given; in {\it d}, the spline is given which has
been used to calculate the \sz~values of the power spectra which were
selected on time.  All points are 64 s averages.
\label{fig:cd}}
\end{center}
\end{figure}

\begin{figure}[]
\begin{center}
\begin{tabular}{c}
\psfig{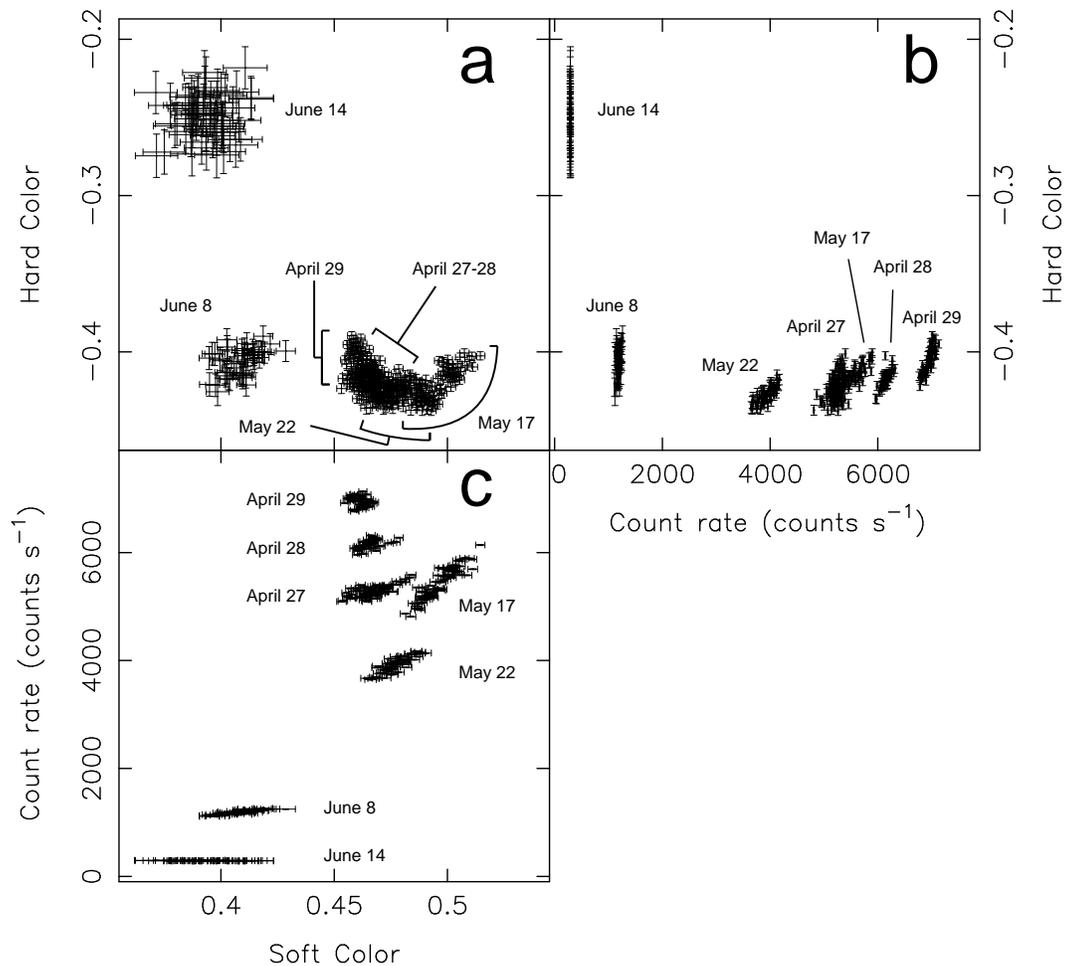}
\end{tabular}
\figcaption{Color-color diagram of XTE J1806--246 created from the
data obtained between 1998 Apr 27 and Jun 14 ({\it a}), excluding the
May 3 data.  In {\it b} the hard color versus 2.0--16.0 keV count rate
is shown (the hard hardness-intensity diagram). In {\it c} the soft
color versus the 2.0--16.0 keV count rate is shown (the soft
hardness-intensity diagram). For the definitions of the colors see
Fig.~\ref{fig:cd}.  The axes in {\it c} are interchanged in order to
allow easier comparison with {\it a}.  All points are 64 s averages.
\label{fig:hid_curved}}
\end{center}
\end{figure}

The CD of all data before 1998 July 1 is presented in Figure
\ref{fig:cd}{\it a}. The data taken on July 1 and 17 are not plotted
because of the very low statistics of the data. In Figure
\ref{fig:cd}{\it a}, different tracks are traced out at different
moments in the outburst. A blow up of the data obtained between April
27 and May 22 is shown in Figure~\ref{fig:cd}{\it b}. During the rise
(the April data) and the decay (the May 17--22 data) the source traced
out a curved branch in the CD (Fig.~\ref{fig:cd}{\it c}). In
Figure~\ref{fig:hid_curved}{\it a} the CD is compared to the hard HID
(Fig.~\ref{fig:hid_curved}{\it b}; the hard color versus the count
rate) and the soft HID (Fig.~\ref{fig:hid_curved}{\it c}; the soft
color versus the count rate, note the axes are interchanged). Clearly
visible from these figures is that although the count rate can be
different by about 2000 counts \pers, the hard and the soft color can
have the same values (this is best visible by comparing the April data
which each other). By comparing the April 27 and the May 17
observation, it is also clear that the soft color can have different
values at the same count rates.  From Figure~\ref{fig:hid_curved} it
can also be seen on which dates the source was at a certain position
in the CD.  During the June observations, the source was in two
distinct places in the CD and HIDs which are separate from the rest of
the data (see Fig.~\ref{fig:cd}{\it a}).

\begin{figure}[]
\begin{center}
\begin{tabular}{c}
\psfig{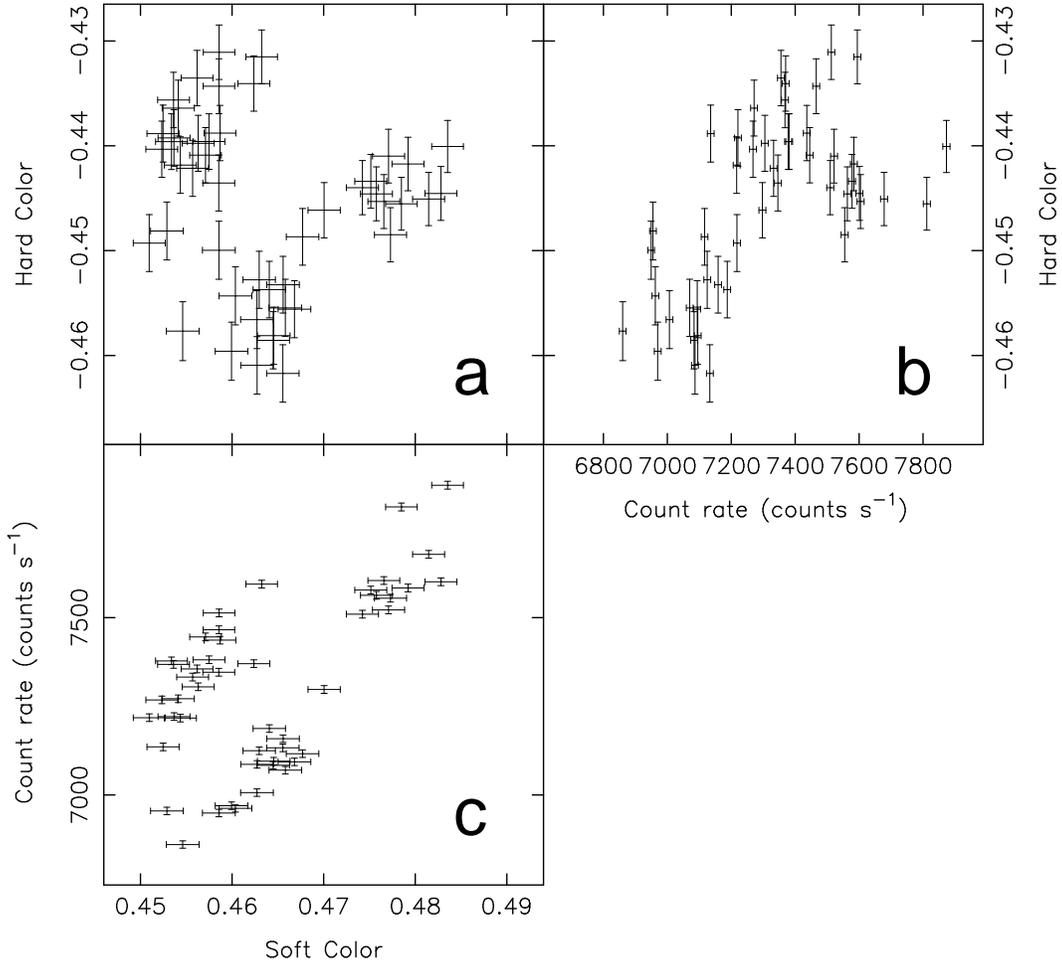}
\end{tabular}
\figcaption{Color-color diagram of XTE J1806--246 created from the
data obtained on May 3 ({\it a}).  In {\it b} the hard color versus
2.0--16.0 keV count rate is shown (the hard hardness-intensity
diagram). In {\it c} the soft color versus the 2.0--1.60 keV count
rate is shown (the soft hardness-intensity diagram). For the
definitions of the colors see Fig.~\ref{fig:cd}.  The axes in {\it c}
are interchanged in order to allow easier comparison with {\it a}.
All points are 64 s averages.
\label{fig:hid_qpo}}
\end{center}
\end{figure}

During the peak of the outburst the source traced out, in a different
part of the CD compared to the rise and decay data, a pattern which we
interpret as a two-branched structure (Fig.~\ref{fig:cd}{\it d}).
When examining the CD at higher (16 seconds) time resolution, it is
clear that at the beginning of the May 3 observation the source was
located on the lower part of the right branch in the CD. The source
then gradually moved via the lower part of the left branch to the
upper left part. At the same time the count rate increased from about
7100 to about 7400 counts \pers~(2.0--16.0 keV).  After a data gap of
about 2000 seconds, due to an Earth occultation of the source, the
count rate had increased to about 7500--7600 counts \pers~(2.0--16.0
keV) and the source was on the upper part of the right branch in the
CD (compare also Fig.~\ref{fig:all_time}{\it a} and {\it b}). It is
impossible to say whether the source jumped to this place in the CD or
moved gradually to it. In the latter case it is also not possible to
say if this gradual motion of the source followed the same track or
followed a different route.  The CD is compared to the HIDs in
Figure~\ref{fig:hid_qpo}.  By comparing {\it a} and {\it c} it can be
seen what the count rate was at a certain position in the CD.

\begin{figure}[]
\begin{center}
\begin{tabular}{c}
\psfig{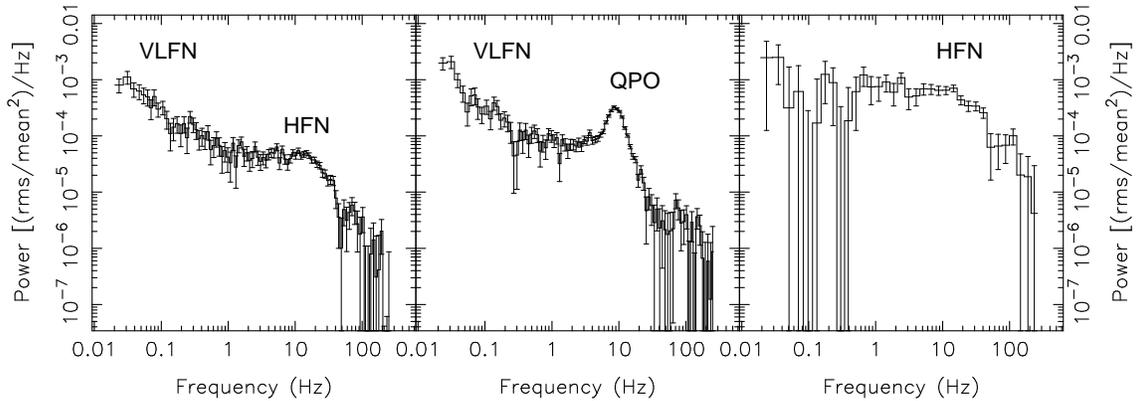}
\end{tabular}
\figcaption{Typical power spectra during the rise and the decay of the
outburst ({\it left}; taken on 1998 April 29), showing the VLFN and
HFN component, the peak of the outburst ({\it middle}; taken on 1998
May 3) showing the QPO, and the last observation in the decay which
has sufficient statistics to detect the noise in the power spectrum
({\it right}; taken on 1998 June 14). The Poisson level has been
subtracted. \label{fig:powerspectra}}
\end{center}
\end{figure}

The source not only traced out different tracks in the CD during the
rise and the decay with respect to the peak of the outburst, but also
the power spectra were remarkably different. During the rise and the
decay two noise components can be distinguished in the power spectrum
(Fig.~\ref{fig:powerspectra} {\it left}): a noise component following
a power law at low frequency (the VLFN) and a broad noise component
which can be represented by a power law with an exponential cutoff
between 5 and 20 Hz (the HFN). The strengths of these noise components
and their other properties during the different observations are
displayed in Table~\ref{obslog}. During the last observation in the
decay for which noise could be detected (1998 June 14; see
Fig.~\ref{fig:powerspectra} {\it right}) the strength of the HFN had
increased to about 14\%. The VLFN could not be detected down to 0.01
Hz (note that Revnivtsev et al. 1999 reported the presence of the VLFN
below 0.01 Hz).  During the last two observations (1998 July 1 and 17)
no noise could be detected, however, the upper limits of typically
30\% rms amplitude on any noise component are not very stringent.
During the peak of the outburst the power spectrum is quite
different. The VLFN is still present but the HFN component is replaced
by a very significant (44$\sigma$) QPO (Fig.~\ref{fig:powerspectra}
{\it middle}) with a frequency of 9.03\pp0.05 Hz, a FWHM of 5.6\pp0.2
Hz, and an rms amplitude of 5.30\pp0.06\%.

\subsection{The QPO}

In order to study this QPO in more detail, we made 64 s FFTs of the
1998 May 3 data and fitted the QPO with a Lorentzian in each
individual power spectrum. In order to correlate the obtained QPO
parameters with the position of the source in the CD, we used the
\sz~parameterization, which has been developed for the Z sources
(\sz~is a measure of the curve length along the track in the CD; see
Wijnands et al. 1997 and Wijnands 1999 and references therein for
detailed descriptions of this procedure). We selected arbitrarily the
normal points where \sz~=~1 and \sz~=~2 (see Fig.~\ref{fig:cd}{\it d}).

\begin{figure}[]
\begin{center}
\begin{tabular}{c}
\psfig{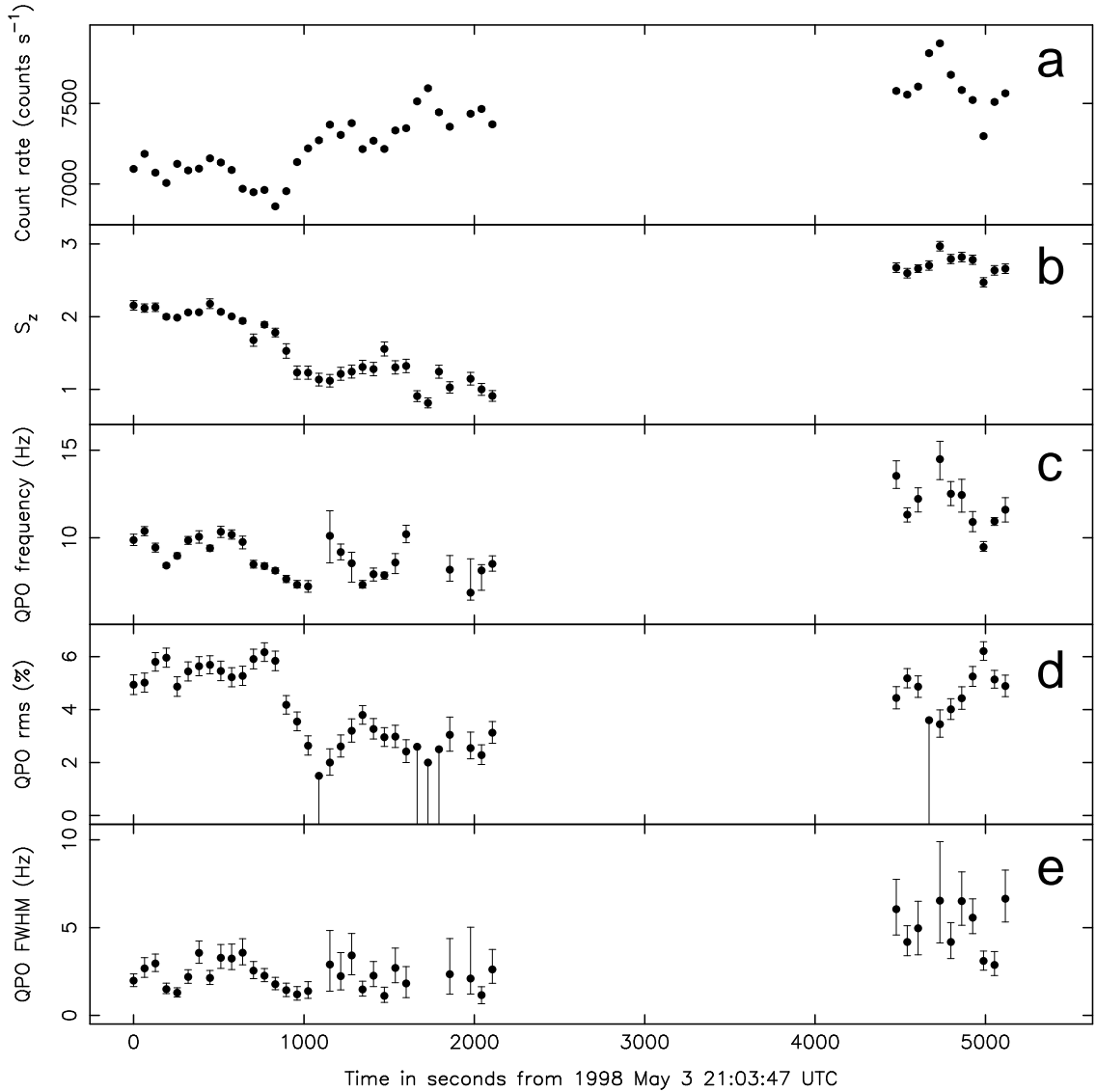}
\end{tabular}
\figcaption{The count rate ({\it a}), \sz~({\it b}), the frequency of
the QPO ({\it c}), the rms amplitude (2--60 keV) of the QPO ({\it d}),
and the FWHM of the QPO ({\it e}) as a function of time from 1998 May
3 21:03:47 UTC. The count rate is in the photon energy band 2.0--16.0
keV and are background subtracted but not deadtime corrected. The gap
between about 2000 and 4500 seconds is due to an Earth occultation of
the source.
\label{fig:all_time}}
\end{center}
\end{figure}

\begin{figure}[]
\begin{center}
\begin{tabular}{c}
\psfig{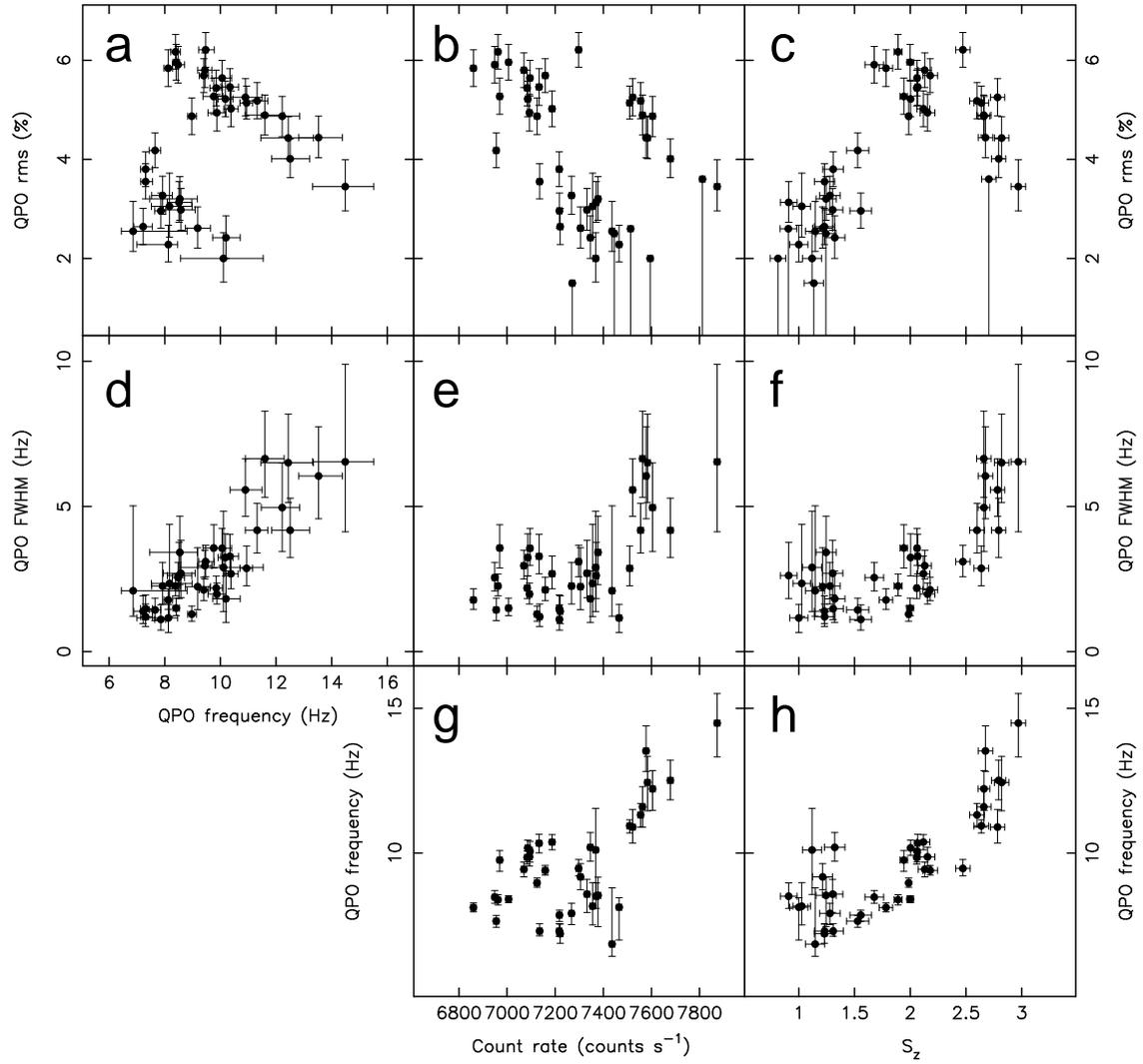}
\end{tabular}
\figcaption{The rms amplitude (2--60 keV) of the QPO as a function of
its frequency ({\it a}), the count rate ({\it b}), and \sz~({\it
c}). The FWHM of the QPO as a function of its frequency ({\it d}), the
count rate ({\it e}), and \sz~({\it f}). The frequency of the QPO as a
function of count rate ({\it g}) and \sz~({\it h}). The count rates
used in these figures are in the energy range 2.0--16.0 keV and are
background subtracted but not deadtime corrected. \label{fig:qpo_all}}
\end{center}
\end{figure}

The behavior of the QPO parameters, the 2.0--16.0 keV count rate, and
the \sz~values as a function of time are presented in
Figure~\ref{fig:all_time}.  The QPO parameters versus the QPO
frequency, the 2.0--16.0 keV count rate, and \sz~are displayed in
Figure~\ref{fig:qpo_all}. From these two figures it is clear that the
QPO parameters (Figs.~\ref{fig:qpo_all}{\it b}, {\it e}, and {\it g})
do not have a strict correlation with the count rate. This is also
apparent when comparing Figure~\ref{fig:all_time}{\it a} with
Figures~\ref{fig:all_time}{\it c}, {\it d}, and {\it e}. The QPO rms
versus its frequency (Fig. ~\ref{fig:qpo_all}{\it a}), shows two
distinct branches. From Figures~\ref{fig:all_time}{\it c} and {\it d}
it can been seen that the transition between these two branches
occurred around 1000 seconds after the beginning of the May 3
observation. After the data gap between 2000 and 4500 seconds (due to
an Earth occultation of the source), the source has returned to the
original branch. From Figure~\ref{fig:qpo_all}{\it c} it is clear that
these two branches correspond to different places in the CD. The
points with the low rms amplitude for the QPO occur below about \sz~=
1.5; the points with the high rms amplitude occur when \sz~$>$
1.5. This differences in QPO behavior are also apparent in
Figures~\ref{fig:qpo_all}{\it f} and {\it h}. Below \sz~= 1.5 the FWHM
and frequency behave erratically. Above \sz~= 1.5 the FWHM and in
particular the QPO frequency increase when \sz~increases. This means
that above \sz~$>$ 1.5 the QPO frequency is well correlated with the
position of the source in the CD. For \sz~$<$ 1.5 the correlation
breaks down.

\begin{figure}[]
\begin{center}
\begin{tabular}{c}
\psfig{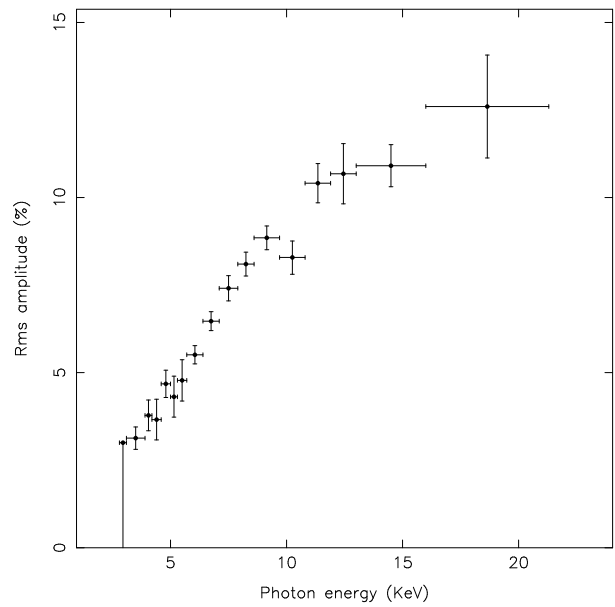}
\end{tabular}
\figcaption{The fractional rms amplitude of the 7--14 Hz QPO versus
the photon energy. \label{fig:qpo_energy}}
\end{center}
\end{figure}

The energy dependence of the QPO is shown in
Figure~\ref{fig:qpo_energy}. There is a clear increase of the rms
amplitude of the QPO with photon energy below $\sim$12 keV. Above 12
keV, the rms amplitude seems to level off. We tried to measure any
time lags in this QPO as a function of energy. The obtained time lags
between the energy bands 2.8--5.3 keV and 5.3--13.0 keV were
consistent with zero ($-0.06$\pp0.4 msec).

\subsection{The noise components during the rise and the decay}

When excluding the May 3 observations in the CD, a curved branch is
present. We investigated the behavior of the noise components as a
function of the position of the source on this branch. We divided the
branch is five parts (see Fig.~\ref{fig:cd}{\it c}). Note that only in
area 3 are data used of the rise {\it and} the decay together. In
areas 1 and 2, only data of the rise is used and in areas 4 and 5 only
data of the decay. During the rise, the source moved first from area 1
via area 2 into area 3 (April 27), and then back via area 2 (April 28)
into area 1 (April 29) (see also Fig.~\ref{fig:hid_curved}).  During
the first observation in the decay (May 17 00:29--00:54), the source
was in area 4. In the subsequent observations, the source moved first
to area 5, and then back via area 4 (May 17 12:42--13:56) into area 3
(May 22).  The motion can be followed when comparing the CD
(Fig.~\ref{fig:hid_curved}{\it a}) with the soft HID
(Fig.~\ref{fig:hid_curved}{\it c}).

The results of the fits in the different areas are summarized in
Table~\ref{noisecdtab}. The VLFN increases in strength and becomes
steeper from the upper left part (area 1) to the upper right part of
the curved branch (area 5), while the HFN decreases in strength. The
index and the cutoff frequency of the HFN do not have a clear
correlation with the position of the source in the CD.  In area 5, the
HFN is not peaked (as is the case in the other areas) and no cutoff
frequency could be determined.

Both noise components have similar relationships with photon energy as
the 7--14 Hz QPO: they increase in strength up to an energy of
$\sim$12 keV above which they remain approximately constant. Although
both the HFN and the QPO peak at roughly the same frequency and have
similar energy dependence of their strengths, it is unclear whether
the peaked HFN evolved into the QPO or was replaced by it.

\subsection{Kilohertz QPOs}

We intensively searched for QPOs between 200 and 1500 Hz. None were
detected with conservative upper limits on the amplitude (2--60 keV;
assuming a FWHM of 150 Hz for the kHz QPOs) of 1\%--2\% rms during the
rise of the outburst, 2.0\% rms during the peak of the outburst,
2.0--3.0\% rms during the decay of the outburst when XTE J1806--246
was on the curved branch, 4.7\% rms on June 8, 12.5\% on June 14, and
around 80\% during the July observations.

\section{Discussion \label{discussion}}

We have analyzed the correlated X-ray spectral and X-ray timing
behavior of the new X-ray transient XTE J1806--246 during its 1998
outburst. During the peak of the outburst we discovered a very
significant (44$\sigma$) QPO near 9 Hz. This QPO was not detected
during the rise and the decay of the outburst, however, a broad peaked
noise component near 10 Hz was. Also, the behavior in the CD was
different between the rise and the decay of the outburst compared with
the peak. During the peak of the outburst a pattern was traced out,
which we interpret as a two branched structure. The QPO frequency
seemed to be correlated with the position of the source in (at least a
part of) the CD: the frequency increased when the source moved from
lower part of the left branch to the upper part of the right
branch. Outside the peak of the outburst, a broad curved branch was
traced out when the count rate was above 3500 counts \pers~(2.0--16.0
keV). Below this count rate only distinct patches in the CD were
formed during the different observations, which were not connect with
each other. This pattern most likely is due to the sparsity of data in
the decay of the outburst. The strength of the HFN increased slightly
when the source moved from the upper right in the CD to the upper
left, while the VLFN strength decreased and it became less steep. The
HFN increased to about 14\% rms amplitude when the count rate had
dropped below 350 counts \pers~(2.0--16.0 keV).

The positional coincidence of XTE J1806--246 with the X-ray burster
SAX J1806.8--2435 makes it likely that they are one and the same
source. This would suggest that XTE J1806--246 contains a neutron
star. Although similar frequency QPOs as the 9 Hz QPO in XTE
J1806--246 have also been discovered in black-hole candidates, they
are more often detected in the neutron star systems (the Z sources,
and on one occasion in an atoll source [Wijnands et al.  1999]). So,
this is in accordance with the idea that XTE J1806--246 contains a
neutron star and not a black-hole.

\subsection{Atoll source versus Z source}

The neutron star LMXBs have been classified into Z sources and atoll
sources (see \S~\ref{intro}). The differences between the Z sources
and the atoll sources have been interpreted (e.g., Hasinger \& van der
Klis 1989) as due to a difference in the accretion rate and in the
strength of the magnetic field of the neutron star. From the higher
intrinsic luminosity of the Z sources compared to that of the atoll
sources, it was suggested that the Z sources accrete near the critical
Eddington accretion rate (e.g., Hasinger \& van der Klis 1989; Smale
1998; Bradshaw, Fomalont, \& Geldzahler 1998) but that most atoll
sources accrete at a significantly lower accretion rate.  The
magnetospheric beat-frequency (MBF) model (Alphar \& Shaham 1985; Lamb
et al. 1985) for the HBOs in the Z sources in combination with the
absence of similar QPOs in the atoll sources, suggested that the
magnetic field strengths of the neutron stars in the Z sources are
higher than those in the atoll sources.  X-ray spectral modeling
(Psaltis et al. 1995) also suggested that the magnetic field strength
in the atoll sources is significantly less than that in the Z sources.
Because the magnetic field strengths of the atoll sources are thought
to be significantly above zero (e.g., Psaltis et al. 1995), the MBF
model predicted that similar QPOs might be observable in the atoll
sources, but at a lower strength than in the Z sources.  With {\it
RXTE} similar frequency QPOs have indeed been found in the atoll
sources, however, it has still to be determined whether the strengths
of these QPOs are significantly less than those of the HBOs in the Z
sources.

The presence of the 5--20 Hz QPO (the N/FBO) in Z sources at their
highest inferred mass accretion rates and the absence of similar QPOs
in the atoll sources, had led to models for this QPO in which the
accretion rate has to be near the Eddington accretion rate in order
for the production mechanism of this QPO to be activated (e.g.,
Fortner, Lamb, \& Miller 1989; Alpar et al. 1992). If these models are
correct, then similar QPOs should also be visible in the atoll sources
when they reach the Eddington mass accretion rate (if they
could). Recently, in the atoll source 4U 1820--30 a 7 Hz QPO was
discovered (Wijnands et al. 1999) at times when this source was
accreting at its highest observed inferred mass accretion
rate. However, this highest observed accretion rate was well below the
Eddington mass accretion rate (Wijnands et al. 1999). If this 7 Hz QPO
is due to the same physical mechanism as the N/FBOs in the Z sources,
then either the production mechanism for this QPO is already activated
at accretion rates much lower than the Eddington accretion rate, or
the accretion rate estimated from the X-ray flux in 4U 1820--30 is
significantly less than the true accretion rate.  In the latter
situation, a significant part of the energy released near the neutron
star must be beamed away from us or has to eventually be emitted not
in X-rays but at other wavelengths or in, e.g., ejecta. It remains to
be seen to what extent the Fortner et al. (1989) model could
accommodate this.

It is interesting to investigate if XTE J1806--246 fits in in the
above described picture, and, if so, how exactly.  During the rise and
the decay of the outburst a curved branch was traced out, which
resembles the banana branch in atoll sources. Also, the behavior of
the noise components with the position of the source on this curved
branch (HFN becoming stronger and the VLFN weaker and less steep when
the source moves from the right to the left in this diagram) is
characteristic of the behavior of the noise components observed in
atoll sources when they move from the upper banana branch to the lower
banana branch. Interestingly, when the count rate drops further in the
decay, distinct patches are formed separated from the curved
branch. The HFN increases to about 14\% rms amplitude with decreasing
count rate and the VLFN becomes undetectable down to 0.01 Hz in the
June 14 observation (the last observation which has sufficient
statistics to detect noise in the power spectrum). The present of such
strong noise and distinct patches in the CD are characteristic of
atoll source behavior when these sources are in the island state (see,
e.g., Hasinger \& van der Klis 1989; M\'endez 1999).  Note, that
similar weak VLFN components, as observed in the June 14 observation
(Revnivtsev et al. 1999), have been observed in other atoll sources
when they were in their island states (Hasinger \& van der Klis 1989).
So, XTE J1806--246 during the rise and the fall of its outburst
displayed all characteristics typical of an atoll source and, if the
source is indeed a neutron star, it should be classified as such.

In the original report of the discovery of the QPO in XTE J1806--246
(Wijnands \& van der Klis 1998) we suggested, that if the 7--14 Hz QPO
in XTE J1806--246 is due to the same physical mechanism as the 7--20
Hz QPOs in the Z sources, XTE J1806--246 might be a Z source. However,
the discovery of the 7 Hz QPO in the atoll source 4U 1820--30
(Wijnands et al. 1999) has since shown that QPOs between 7--20 Hz are
in neutron star systems not exclusively found in the Z sources. The
correlation of the frequency of the QPO with the position of the
source in part of the CD is very similar to the behavior of the NBO
when the Z sources move from the lower NB onto the lower FB. Such
correlated behavior has not yet been seen in other atoll
sources. However, the lack of correlation in another part of the CD is
different. We conclude that if XTE J1806--246 is a neutron star it is
most likely the second example of an atoll source that at high
luminosity begins to show a NBO-like QPO. It then is the first example
of an atoll source simultaneously exhibiting a NB/FB like pattern in
the CD. (A related case may be Cir X-1, which sometimes shows atoll
characteristics [Oosterbroek et al. 1995] and at other times what
appears to be a Z-track and HBO- and NBO-like QPOs [Shirey et
al. 1998, 1999]).

It is possible that XTE J1806--246 during the peak of the outburst
reaches near Eddington mass accretion rates, as is required in the
Fortner et al. (1989) model for the NBO.  The observed absorbed flux
between 2 and 30 keV during the peak of the outburst is about
1.9$\times10^{-8}$ ergs cm$^{-2}$ \pers. The source is located in the
direction to the galactic bulge ($l \sim 6.1$, $b \sim
-1.9$). Assuming a distance of 8 kpc, the intrinsic luminosity of the
source is 1.5$\times10^{38}$ ergs \pers, which is close to the
Eddington accretion limit for a 1.4 solar mass neutron star. However,
during the rise and the decay the observed X-ray count rate and flux
($\sim1.8\times10^{-8}$ ergs cm$^{-2}$ \pers) are very close to those
observed during the peak of the outburst, suggesting that also at
those epochs XTE J1806--246 was accreting near the Eddington accretion
limit. It is unclear why only during the peak of the outburst XTE
J1806--246 showed the 7--14 Hz QPO; perhaps this is related to a
difference in the accretion flow between persistent and transient
sources caused by the large changes in the mass accretion rate in the
latter.

\subsection{Other types of QPOs?}

Both the Z sources and the atoll sources exhibit, at their lowest
observed inferred mass accretion rates, kHz QPOs and QPOs between
15--70 Hz.  Therefore, we looked in particular at the data obtained
during the lowest count rates for XTE J1806--246 to search for similar
QPOs. No significant kHz QPOs were detected but with upper limits
which are not inconsistent with the derived values of the kHz QPOs in
the Z sources and in some atoll sources.  Interestingly, when we
combine the power spectra obtained for areas 1 and 2 in
Figure~\ref{fig:cd}{\it c}, we detect a 4.8$\sigma$ QPO with a
frequency of 70.6\pp1.6 Hz, a FWHM of 12.6$^{+4.6}_{-3.2}$ Hz, and an
rms amplitude of 1.2\%\pp0.1\%. This QPO is also marginally visible in
Fig.~\ref{fig:powerspectra} {\it left}.  However, when we include the
Poisson level as an unknown parameter in the fit the significance of
this QPO drops to about 3.5$\sigma$. Taking into account the number of
trials involved in our search method this QPO needs
confirmation. However, the presence of such a $\sim$70 Hz QPO is very
similar to similar frequency QPOs observed in atoll sources in their
lower banana branch and to the HBOs observed in the Z sources on their
HB.

\subsection{Conclusion}

The behavior observed for XTE J1806--246 during the 1998 outburst is
consistent with that of an atoll source which at the peak of the
outburst reached a X-ray luminosity level where the NBO mechanism
switched on. This makes the source the second atoll source, after 4U
1820--30 (Wijnands et al. 1999), that exhibits QPOs with frequencies
near 7--9 Hz. This source can be very useful to study the similarities
and the differences between Z sources (including Cir X-1, which is
probable a Z source; Shirey et al. 1998; 1999) and atoll sources,
allowing us to get a better insight in the processes at work in the
inner part of the accretion disk.

\acknowledgments

This work was supported in part by the Netherlands Foundation for
Research in Astronomy (ASTRON) grant 781-76-017, by the Netherlands
Researchschool for Astronomy (NOVA), and the NWO Spinoza grant 08-0 to
E. P. J. van den Heuvel.  This research has made use of data obtained
through the High Energy Astrophysics Science Archive Research Center
Online Service, provided by the NASA/Goddard Space Flight Center. We
thank the anonymous referee for his helpful comments on the paper.

\clearpage

\clearpage

\begin{deluxetable}{lllllllllll}
\tablecolumns{11}
\tiny
\tablewidth{0pt}
\tablecaption{Log of the observations and the fit parameters of the noise components during those observations\label{obslog}}
\tablehead{
\colhead{}     & \colhead{} &\colhead{}     & \colhead {}     & \colhead{}            & \multicolumn{2}{c}{VLFN}                       & & \multicolumn{3}{c}{HFN}                                    \\ \cline{6-7}\cline{9-11}
Obs. ID        & Date       & Start -- End  & On source time  & Count rate range$^a$  & Rms$^b$               & Index                  & & Rms$^c$              &  Index               & Cutoff freq. \\         
\colhead{}     & (n 1998)   &  (UT)         &  (ksec)         & (counts \pers)        & (\%)                  &                        & & (\%)                 &                      & (Hz)         \\}        
\startdata
30412-01-01-00 & Apr 27     & 19:31--21:44  & 5.2             & 5000--5630 (5290)     & 1.3$^{+0.2}_{-0.1}$   & 0.8\pp0.1              & & 3.7\pp0.1            & $-$1.2\pp0.2           & 7.2\pp0.7 \\            
30412-01-02-00 & Apr 28     & 18:59--20:30  & 2.6             & 5895--6405 (6150)     & 1.3\pp0.1             & 1.0\pp0.1              & & 3.3\pp0.1            & $-$1.8\pp0.3           & 6.0\pp0.8 \\            
30412-01-03-00 & Apr 29     & 20:42--21:38  & 3.3             & 6700--7195 (6950)     & 1.26\pp0.06           & 0.77$^{+0.08}_{-0.10}$ & & 3.3\pp0.2            & $-$1.2$^{+0.3}_{-0.4}$ & 9.1\pp1.5 \\            
30412-01-04-00 & May 03     & 21:01--22:32  & 2.9             & 6810--8045 (7315)     &  1.3\pp0.1            & 0.99\pp0.09            & & --                   & --                   & --        \\            
30412-01-05-01 & May 17     & 00:29--00:54  & 1.5             & 4855--5585 (5215)     & 3.0$^{+0.5}_{-0.4}$   & 1.29$^{+0.10}_{-0.09}$ & & 3.1\pp0.2            & $-$1.9$^{+0.5}_{-0.6}$ & 5.5$^{+1.3}_{-1.1}$ \\  
30412-01-05-00 & May 17     & 12:42--13:56  & 3.4             & 4720--6255 (5515)     & 4.2\pp0.3             & 1.42\pp0.05            & & 1.4\pp0.2            & $-$2.0\pp1.4           & 3.5$^{+3.5}_{-1.3}$ \\  
30412-01-06-00 & May 22     & 11:06--12:07  & 3.5             & 3495--4310 (3930)     & 3.8 $^{+0.4}_{0.3}$   & 1.38$^{+0.07}_{-0.06}$ & & 3.42\pp0.09          & $-$2.6\pp0.4           & 4.0\pp0.6 \\            
30412-01-07-00 & Jun 08     & 07:59--09:30  & 2.9             & 1075--1285 (1190)     & 7.8$^{+2.7}_{-1.5}$   & 2.0\pp0.2              & & 5.7\pp0.4            & $-$1.3$^{+0.4}_{-0.5}$ & 12.6$^{+3.8}_{-2.8}$ \\ 
30412-01-08-00 & Jun 14     & 20:54--22:02  & 3.6             &  271--313  (290)      & $<$5.3                & 2$^d$                  & & 14.3$^{+0.7}_{-0.5}$ & $-$0.2\pp0.1           & 19.0$^{+4.2}_{-3.2}$ \\ 
30412-01-09-00 & Jul 01     & 19:45--20:23  & 0.9             &   15--29    (22)      &                       &                        & & $<$30                &                      &                      \\
30412-01-10-00 & Jul 17     & 19:50--20:24  & 1.7             &   13--28    (20)      &                       &                        & & $<$29                &                      &                      \\
\enddata 
\tablenotetext{a}{In the photon energy range 2.0--16.0 keV. The values between brackets are the averaged count rates}
\tablenotetext{b}{Integrated over 0.001--1.0 Hz; 2--60 keV}
\tablenotetext{c}{Integrated over 1.0--100.0 Hz; 2--60 keV}
\tablenotetext{d}{Parameter fixed}
\end{deluxetable}

\begin{deluxetable}{lllllll}
\tablecolumns{7}
\tablewidth{0pt}
\tablecaption{Noise parameters versus position in the CD\label{noisecdtab}}
\tablehead{
        & \multicolumn{2}{c}{VLFN}                          & & \multicolumn{3}{c}{HFN}                 \\\cline{2-3}\cline{5-7}
CD area$^a$ & Rms$^b$               & Index                  & & Rms$^c$              &  Index               & Cutoff freq. \\
Number  & (\%)                  &                        & & (\%)                 &                      & (Hz)         \\}
\startdata
1       & 1.27\pp0.05           & 0.77$^{+0.06}_{-0.05}$ & & 3.4\pp0.1            & $-$1.3\pp0.2           & 8.6$^{+1.1}_{-1.0}$ \\
2       & 1.6\pp0.1             & 1.09\pp0.08            & & 3.77\pp0.08          & $-$1.1\pp0.2           & 7.9\pp0.7\\
3       & 3.1$^{+0.3}_{-0.2}$   & 1.32\pp0.06            & & 3.44\pp0.08          & $-$1.8\pp0.3           & 5.0\pp0.6\\
4       & 2.9$^{+0.4}_{-0.2}$   & 1.29\pp0.07            & & 2.8\pp0.1            &  $-$1.6\pp0.4          & 6.2$^{+1.2}_{-1.1}$ \\
5       & 5.2$^{+0.7}_{-0.5}$   & 1.56\pp0.09            & & 1.9\pp0.2            &  0.6\pp0.1           & $\infty^d$ \\
\enddata 
\tablenotetext{a}{For the selections see Fig.~\ref{fig:cd}{\it c}}
\tablenotetext{b}{Integrated over 0.001--1.0 Hz; 2--60 keV}
\tablenotetext{c}{Integrated over 1.0--100.0 Hz; 2--60 keV}
\tablenotetext{d}{Parameter fixed}
\end{deluxetable}

\end{document}